\documentclass[12pt]{article}
\usepackage{epsfig,amssymb}
\usepackage{latexsym}

\newcommand{\bq}{\begin{equation}}
\newcommand{\eq}{\end{equation}} \newcommand{\bqa}{\begin{eqnarray}}
\newcommand{\eqa}{\end{eqnarray}} \newcommand{\ben}{\begin{enumerate}}
\newcommand{\een}{\end{enumerate}}
\newcommand{\bc}{\begin{center}}
\newcommand{\ec}{\end{center}} \newcommand{\bqb}{\begin{eqnarray*}}
\newcommand{\eqb}{\end{eqnarray*}}

\def\pr#1#2#3{ Phys. Rev. ${\bf{#1}}$, #2 (#3)}

\def\pl#1#2#3{ Phys. Lett. ${\bf{#1}}$, #2 (#3)}

\def\np#1#2#3{ Nucl. Phys. ${\bf{#1}}$, #2 (#3)}
\def\zp#1#2#3{ Z. f. Phys. ${\bf{#1}}$, #2 (#3)}

\def\epj#1#2#3{ Eur. Phys. J. ${\bf{#1}}$, #2 (#3)}

\def\ie{{\it i.e. }}
\def\eg{{\it e.g. }}

\global\nulldelimiterspace = 0pt

\begin{document}
\pagenumbering{arabic}
\thispagestyle{empty}
\def\thefootnote{\fnsymbol{footnote}}
\setcounter{footnote}{1}

\begin{flushright}
January 2005\\
PM/05-1\\
hep-ph/0501046\\
corrected version\\

 \end{flushright}
\vspace{2cm}
\begin{center}
{\Large\bf About helicity conservation in gauge boson scattering
at high energy\footnote{Work supported by the European Union
under contract HPRN-CT-2000-00149.}}.  \vspace{1.cm}  \\
{\large G.J. Gounaris$^a$ and F.M. Renard$^b$}\\
\vspace{0.2cm}
$^a$Department of Theoretical Physics, Aristotle
University of Thessaloniki,\\
Gr-54124, Thessaloniki, Greece.\\
\vspace{0.2cm}
$^b$Laboratoire de Physique Th\'{e}orique et Astroparticules,
UMR 5207\\
Universit\'{e} Montpellier II,
 F-34095 Montpellier Cedex 5.

\vspace*{1.cm}

{\bf Abstract}
\end{center}
\vspace*{-0.4cm}
We remark that the high energy gauge boson scattering processes
involving two-body initial and final states, satisfy certain  
selection rules described as helicity conservation of the
\underline{gauge boson} amplitudes (GBHC). These rules are valid 
at Born level, as well as at the level of the leading and 
sub-leading 1-loop logarithmic corrections, in both the
Standard Model (SM) and  the Minimal Supersymmetric Standard Model
(MSSM). A "fermionic equivalence" theorem is also proved,
which suggests that  GBHC is valid at all orders in MSSM at 
sufficiently high energies, where the mass suppressed
contributions are neglected.

\vspace{0.5cm}

\def\thefootnote{\arabic{footnote}}
\setcounter{footnote}{0}
\clearpage

Many people may have noticed that at high energy where masses are
 neglected, two-body processes involving \underline{transverse} gauge bosons
($V=$ gluon, photon, $Z,W^{\pm}$) satisfy certain
selection rules implying asymptotic helicity conservation in the s-channel.
This can easily be seen at Born level in either the Standard Model (SM)
or its renormalizable  SUSY extensions;
\eg the Minimal Supersymmetric Standard Model (MSSM).
For example, considering
the processes $V_{\lambda_V}+V'_{\lambda_{V'}}
\to A_{\lambda_A}+A_{\lambda_{A'}}'$,
and computing the diagrams of Fig.1 corresponding to $A,A'$ beings scalars,
one observes that   the high energy helicity amplitudes
$F_{\lambda_V \lambda_{V'}\lambda_A\lambda_{A'}}$
vanish for    $\lambda_V=\lambda_{V'}$; while for the fermion
production  case of Fig.2, the vanishing of the high energy amplitudes
is guaranteed whenever either of the  relations
$\lambda_V=\lambda_{V'}$ or $\lambda_A=\lambda_{A'}$ is satisfied.
Correspondingly,  the
amplitudes for  the crossed process
$V_{\lambda_V}+A_{\lambda_A} \to V'_{\lambda_{V'}}+A_{\lambda_A'}'$
vanish when $\lambda_V=-\lambda_{V'}$ for the case of Fig.1; or when
either of the relations $\lambda_V=-\lambda_{V'}$ or
$\lambda_A=-\lambda_{A'}$ is   satisfied for  the fermion case of Fig.2.  \par

Similar asymptotic rules  also exist  for the purely gauge helicity amplitudes
$F_{\lambda_1\lambda_2\lambda_3\lambda_4}$ of the processes
$V_{1\lambda_1}V_{2\lambda_2} \to V_{3\lambda_3}V_{4\lambda_4}$
involving four gauge bosons. Thus, it has been observed
in \cite{GLRWW} that these amplitudes satisfy asymptotically
\bqa
&&F_{+++-}=F_{++-+}=F_{+-++}=F_{-+++}=F_{---+}\nonumber\\
&&=F_{--+-}=F_{-+--}
=F_{+---}=F_{++--}=F_{--++}=0~~ \label{gauge-Born}
\eqa
  at the Born level, in either SM or MSSM.
Consequently, only the helicity amplitudes satisfying
$\lambda_1+\lambda_2=\lambda_3+\lambda_4$ can survive
asymptotically,  at this  level.

These  properties of  gauge boson helicity
conservation (GBHC), are  a priori different
and complementary to the well-known fermion helicity
conservation in processes involving external fermions.
The later is an  essentially  kinematical
consequence  of the fermionic  vertices  in  SM or MSSM,  valid
at a diagram by diagram basis,  provided that the energy
is sufficient high, so that all masses
can be neglected\footnote{See below the discussion of the effects
of Yukawa couplings.}.

GBHC though, referring specifically to the external gauge boson helicities,
is  more subtle.
Contrary to the fermionic case,
detail cancellation among the contributions of various diagrams
must take place, before GBHC  is established.
This can  be seen from the Born processes  described
by Figs.1 or 2, where  the asymptotic vanishing  of the helicity amplitudes
for $\lambda_V=\lambda_{V'}$ is established through
the  occurrence
of "large gauge cancellations" among the $Vff$ and $VVV$
vertices; or  among the
$Vss$, $VVV$ and $VVss$ vertices, with $s$ describing
generic scalar particles. It should also be emphasized that
such  cancellations are only
realized when the minimal gauge couplings, characterizing
the renormalizable gauge theories, are used. They  would be violated if \eg
higher dimensional operators  are inserted the theory, even
though $SU(3)\times SU(2)\times U(1)$ gauge symmetry is  still
 respected \cite{GLRWW}. Renormalizability of the theory
is therefore  crucial, for these
rules to be valid\footnote{The simplest illustration is the scalar
coupling of the type $\phi F^{\mu\nu}F_{\mu\nu}$. It
is perfectly gauge invariant, but if used in a
scalar exchange diagram, it violates the above rules.
Another simple example is the "anomalous"
quadruple coupling \cite{Schi, GLRWW}. The complete list of such
anomalous gauge invariant couplings can be found in \cite{GR}.}.\\

Sofar we have only considered  tree  diagrams, and one may
wonder whether these high energy helicity conservation
 properties remain true  beyond the Born approximation.
Indeed, for processes receiving  a Born contribution,
one can immediately check that these properties   remain true
at the level of the 1-loop  leading $\ln^2 s$ and subleading
$\ln s$ logarithmic corrections,   according to the theory developed
in  \cite{BRV,DP, QCD}.
This we have also checked explicitly for
$e^-e^+\to \gamma \gamma ~,~ZZ~,~ \gamma Z$ using the
complete 1-loop results of \cite{LC-VV}, and for  $e^+e^-\to W^+W^-$
using \cite{WW}.\par

We have also looked at the process
$\gamma \gamma \to \gamma \gamma $ \cite{Laz-4gamma},
$\gamma \gamma \to ZZ $  \cite{Laz-ZZ} and
$\gamma \gamma \to \gamma Z $ \cite{Laz-gamZ}, where
there is no Born term and the high energy 1-loop behavior
is known. The validity of GBHC for the leading and sub-leading
 logarithmic terms is again observed in both SM and  MSSM.
However, at the level of the sub-sub-leading (constant)
1-loop contributions, GBHC is generally
violated within  SM, but it is still preserved in MSSM. \\

Motivated by this observation and the surprising
analogy between the fermionic  helicity conservation and GBHC,
we have looked at its justification,  on the basis of    supersymmetric
invariance  and renormalizability. The aim of the present paper
is to release this justification.\par

We work in the framework of the
exact supersymmetric limit of MSSM, assuming in addition that
 the Higgs-bilinear $\mu$-term of the superpotential is
 also vanishing.  In such a theory, all particles are massless, and the
 electroweak gauge symmetry  is not broken.
We denote the leptons and quarks by
  the chiral  spin=1/2  fields $(\psi_L,\psi_R)$,
  the sleptons and squarks by   the corresponding  scalar fields
$(\tilde \psi_L,\tilde \psi_R)$, the gauge bosons by $V_j^\mu$, their
gaugino partners by $\chi_j=\chi_{jL}+\chi_{jR}$, the
higgsino doublets by  $\tilde H_{(1,2)L}$, and  the corresponding
Higgs doublets  by  $H_{(1,2)L}$. The later  include also the
Goldstone bosons.

In fact, since all particles are massless in this theory,
 the notation of the fermionic fields
 may be further simplified by denoting them as
  $(\psi_\lambda,~\chi_\lambda)$,
 with $\lambda$ being the helicity of the particle the field  absorbs.
 The  corresponding scalar  fields may also  be defined by this
 helicity and  written as  $\tilde \psi_\lambda$;
 in fact it is advantageous
to think of this scalar field as carrying a "formal  helicity" $2\lambda$.
The same definition applies also to higgsino and Higgs
fields.
In this  massless theory, all  purely scalar  self interactions
consist of 4-leg-vertices arising either from the F-terms generated by the
superpotential, or from D-terms. In each of these vertices the  total
"formal helicity" defined above is conserved.

The sum of fermion helicity and
 "formal  helicity" of the scalar fields,
 is also conserved in all gaugino-fermion-sfermion
 and gaugino-higgsino-Higgs MSSM vertices. Thus,
 \eg a massless quark of a definite
 helicity can be transformed to an opposite gaugino helicity,
emitting at the same time a scalar field, that remembers it;
so that the sum of the  fermion-helicity  and the
"formal  helicity" is conserved at each vertex separately.

The fermion helicity
in each of the gauge-fermion  vertices, is of course also conserved,
for all kinds of  fermions, including gauginos and
higgsinos. In this respect, we   think of the
massless  gauge bosons of our theory as carrying
vanishing "formal helicity", and claim that all gauge-fermion vertices
also conserve the sum of fermion and  formal helicities.

It might be useful to think of this  conservation  of the sum of
fermion and formal helicities, as a new global $U(1)$ symmetry
respected by all vertices in our framework, except
 the fermion vertices induced by the
Yukava terms in the superpotential.

However, if we restrict to
 processes determined by diagrams in which the Yukawa terms can only
  appear in hermitian conjugate  pairs, then
  this overall generalized  helicity conservation rule will not be affected.
Since we only consider  two-body scattering amplitudes,
this is achieved \eg by restricting to processes
 involving  an even  number of external
transverse gauge bosons, and/or  an even number of external
gauginos.
In such amplitudes, the number of external Higgs fields,
as well as the number of external Higgsinos, are also  always
even. These are in fact the processes which
constitute our main interest.\\

With these definitions, it is straightforward to check
helicity conservation
for any 2-fermion to 2-fermion process at high energy,
when all masses are neglected. More explicitly, in any
allowed  such  process,
 the helicities of the incoming and outgoing
particles in an amplitude which is not forced to
vanish asymptotically,
 should satisfy
\bq
F(f_\lambda f_{\lambda'}' \to f_\mu f_{\mu'}') ~~~
  \Leftrightarrow  ~~~ \lambda+\lambda'=\mu+\mu' ~~, \label{4f-hc}
\eq
to all orders in
our framework. We emphasize that this result is valid
 separately for each contributing diagram,
 independently of  the nature
of the fermions involved; \ie whether some or all of them
are quarks or leptons or their antiparticles, or gauginos, or higgsinos.

The same result (\ref{4f-hc}) remains true, if two of the fermions
(irrespective of whether they are in- or out-going)
are replaced by scalars.
In this case of course, the helicities for the scalar
particles actually refer to
their "formal helicities" defined above. Since these are $\pm 1$  though,
while the fermionic ones are half-integers,
it is  immediately seen that the only  relevant amplitudes
which may be asymptotically non-vanishing, have the structure
\bq
F(f_\lambda s  \to f_{\lambda}' s') ~~~~{\rm or}~~~~
F(f_\lambda f'_{-\lambda}  \to s  s') ~~,  \label{2f-2s-hc}
\eq
where  $(s,s')$ denote any kind of
scalars\footnote{Including of course also
the  Goldstone bosons.}, and $(f,f')$ are
fermions with their helicities indicated as indices in (\ref{2f-2s-hc}).

It is important to realize   that  (\ref{4f-hc}, \ref{2f-2s-hc})
 imply  conservation of \underline{physical helicities}
 at asymptotic energies, for any processes involving only
 external fermions and/or   scalars.
 The physical helicities of all
scalars are, of course, vanishing.\\

For proving GBHC for the \underline{physical helicities} of
the transverse gauge bosons, we just  rely  upon  the validity
of (\ref{4f-hc}, \ref{2f-2s-hc}),
and the supersymmetric transformation properties of the
external fields\footnote{The notion of "formal helicity" is not needed for this.}.
For simplicity we start  from the 2-fermion to 2-fermion
amplitudes in (\ref{4f-hc}),
for the case where
all incoming and outgoing fermions describe gauginos. We then remark that
the supersymmetric transformation for the gaugino fields is
\bq
\delta \chi^j =\frac{1}{2}\sigma^{\mu\nu} F^j_{\mu\nu}\gamma_5
\epsilon -D^j \epsilon ~~,
\label{SUSY-tran1}
\eq
where $j$ is the gaugino  group index,  $F^j_{\mu\nu}$ and $D^j$ are
the corresponding gauge-strength and auxiliary fields,
and $\epsilon$ is the usual SUSY Majorana constant  \cite{SUSY}.
This implies that a massless incoming gaugino state of helicity
$\mu$ and momentum $p$ along the $\hat z$-axis,
transforms completely  into a massless gauge state
with helicity $\lambda$ and the same momentum and gauge
quantum numbers. The explicit result is\footnote{The derivation of
this relation only involves the standard  algebra
for the  massless  fermionic  and gauge states, for  the aforementioned
momenta and helicities.}
\bq
\delta \chi_\mu = \delta \left ( \frac{(1+2\mu \gamma_5)}{2}\chi^j
 \right )=
\frac{(1+2\mu \lambda )}{2} ~\frac{i p}{\sqrt{2}}(1+\lambda \gamma_5)
(i\lambda \sigma^{23}+\sigma^{13})\epsilon ~~. \label{SUSY-tran2}
\eq
The crucial term in  (\ref{SUSY-tran2}) is  the  factor
$(1+2\mu \lambda)$ on the r.h.s,
which guarantees   that the helicities of the
transverse  gauge bosons generated under a SUSY transformation,
  will always have   the same signs
as those  of the initial  gauginos\footnote{The D-term in
(\ref{SUSY-tran1}), being always a product of 4 fields
in an unbroken SUSY theory, gives no contribution to the
single particle projection in (\ref{SUSY-tran2}). }.
Thus, any asymptotic
helicity structure of the 2-gaugino to 2-gaugino process,
will be transformed into a
2-gauge to 2-gauge process having the same structure.
Starting therefore from
(\ref{4f-hc}) applied to gauginos, we conclude
that the physical helicities of the   asymptotically
non-vanishing 2-transverse
gauge to 2-transverse
gauge amplitudes,  satisfy
\bq
F(V_\lambda V_{\lambda'}' \to V_\mu V_{\mu'}')
~~~\Leftrightarrow
~~~ \lambda+\lambda'=\mu+\mu' ~~, \label{4V-hc}
\eq
to all orders in our framework.

This procedure can be straightforwardly extended to amplitudes
involving any even
number of gauginos. Thus,  the only
 asymptotically non-vanishing amplitudes involving
 two transverse gauge bosons should have the helicity structure
 \bq
 F(V_\lambda f_\mu \to  V'_\lambda f'_\mu) ~~ , ~~
F( V_\lambda V'_{-\lambda} \to f_\mu f'_{-\mu})~~,~~
 F( V_\lambda s \to  V'_\lambda s') ~~ , ~~
 F(V_\lambda V'_{-\lambda} \to s s')~~, \label{2V-hc}
\eq
with  $(f,f')$ and $(s,s')$ being  fermions and  scalars
respectively; with the  appropriate quantum numbers of course, so that
the process is allowed.\\

In the above study we have proved the
"physical helicity" conservation rules
(\ref{4f-hc}, \ref{2f-2s-hc}, \ref{4V-hc}, \ref{2V-hc}),
in an exactly supersymmetric theory,
where all particles are massless and electroweak symmetry (EW) is not broken.
Longitudinal gauge bosons do not exist in this theory, but the
Goldstone boson (Higgs) fields do  appear,
among the scalar  external states of  (\ref{2f-2s-hc},  \ref{2V-hc}).

After EW breaking and masses are  generated,
 (\ref{4f-hc}, \ref{2f-2s-hc}, \ref{4V-hc}, \ref{2V-hc}) will of course
remain asymptotically true for transverse gauge bosons.
At the same time,  the  equivalence theorem,   guarantees that the
external Goldstone bosons  may readily be replaced
by longitudinal gauge bosons in these amplitudes  \cite{equiv}.
Thus, if \eg the scalars in the last of the amplitudes  (\ref{2V-hc}) are Goldstone
bosons, then the equivalence theorem guarantees  also the
existence of  the asymptotic amplitudes
\[
F(V_\lambda V'_{-\lambda} \to V''_0 V'''_0) ~~.
\]

Doing such replacements, in all possible ways, it is easy to see that the
complete set of the asymptotically allowed gauge-involving amplitudes
is again described by\footnote{The purely Goldstone
four-body amplitude will, of course also be needed here.}
(\ref{4V-hc}, \ref{2V-hc}), with  the
vector bosons helicities now allowed to acquire vanishing values, while
$(s,s')$  are now  interpreted as sfermions or   physical Higgs particles
only\footnote{In principle, it could even
be possible to have asymptotically non vanishing amplitudes of the form
$V_0 s\to s' s''$, where the vector boson is longitudinal.
Conservation of other quantum numbers like \eg
CP,  forbids the appearance of such terms in MSSM. }.
Eqs.(\ref{4f-hc}, \ref{2f-2s-hc}) will of course also remain true,
under  this  interpretation.\\

The above proof of  "fermionic equivalence" assumes that  SUSY is indeed
realized in Nature at a moderate scale, such that the corresponding
selection rules can be observed at high energy. In such a case in
fact, eqs.(\ref{4f-hc}, \ref{2f-2s-hc}, \ref{4V-hc}, \ref{2V-hc})
can be extended to any two-body
process which is not determined by diagrams of odd order in the
Yukava couplings. Thus,  the
asymptotically non-vanishing  amplitudes should satisfy
\bq
F(a_{\lambda_1}b_{\lambda_2}
 \to c_{\lambda_3} d_{\lambda_4})
~~~\Leftrightarrow
~~~ \lambda_1+\lambda_2=\lambda_3 +\lambda_4 ~~, \label{4body-hc}
\eq
to all orders in $\alpha$, for any kind of particles $(a,b,c,d)$
with  physical helicities
$(\lambda_1, \lambda_2,\lambda_3, \lambda_4)$, provided the process is
of even order on the Yukawa couplings,
and it is of course allowed.
As already mentioned, a sufficient condition for this is that
the process involves an even number of transverse  gauge and
an even number of  gaugino states.
If both initial particles have spin $1/2$, and the final are gauge
or scalar bosons, (or vice versa),
the helicity constraint  in  (\ref{4body-hc})
is further restricted as
$\lambda_1+\lambda_2=\lambda_3 +\lambda_4=0$; while if one of the
particles in each of the initial and final state has spin $1/2$,
and the other is boson, helicity is conserved separately for the
fermions and  the bosons of the process; compare (\ref{2V-hc}).

In case SUSY would not be realized at a moderate scale, or not realized
at all, then SM will provide the appropriate framework.
In this framework, GBHC would remain valid only at the Born
approximation, including the leading and sub-leading
1-loop logarithmic corrections\footnote{Occasionally it may be possible
to extend this rule to non asymptotic energies also.
As an example we mention
the tree level observation in \cite{Parke}, that
the projections of the $t$ and $\bar t$ spins along the
"off diagonal axis" in the $e^-e^+ \to t \bar t $ c.m. frame,
 must be equal for  any energy. This "off-diagonal"
axis coincides asymptotically with the $t-\bar t$ helicity axis.}.
Depending on the process,
it may be broken at the sub-sub-leading (constant) level, though.
We have already mentioned that
this is the case in 2-gauge boson to 2-gauge boson processes.
Specific studies of other processes  should be
done in order to see if this is a general feature, \ie if indeed
 there is a residual GBHC-violating term in SM,
which is only  cancelled when  the supersymmetric partner
contributions are added.
A priori, there could also be cases in which the
sub-sub-leading terms cancel
separately in SM and in SUSY contributions.\\

Incidently one should also mention that the cancellation of the
GBH-violating amplitudes
leads to a remarkable simplification of the actual
theoretical description of the processes; about half of the
helicity amplitudes disappear and the expressions of
the remaining ones are noticeably simplified.

Theoretically, GBHC looks like an appealing simple rule.
Experimentally, it may be possible to check it  at LHC or ILC,
by looking at processes involving gluons, photons, $Z$
or $W$'s in processes like
\bqa
 &&  q\bar q \to gg~,~ g\gamma ~,~ gZ~,~ gW~,~ \gamma\gamma ~,~
\gamma Z~,~ ZZ~,~ W^+W^- ~,~ \gamma W~,~  ZW~~, \nonumber  \\
&&   gq \to gq ~,~ \gamma q ~,~ Zq ~,~ Wq ~, \nonumber \\
&&  gg \to gg ~,~ q\bar q ~, \nonumber \\
&&  e^+e^-\to \gamma\gamma ~,~ \gamma Z~,~ZZ~,W^+W^-~ ,\nonumber \\
 && \gamma e\to \gamma e ~,~ Ze~,~ W\nu ~,~ \nonumber \\
 && \gamma \gamma \to  f\bar f ~,~ \gamma\gamma ~,~
\gamma Z~,~ ZZ~,~ W^+W^-~,~ \nonumber
\eqa
as well as processes involving external  supersymmetric
particles, like \eg
$ gg\to \tilde g\tilde g~,~ \tilde q\bar {\tilde q}$ and
$\gamma\gamma \to \tilde f \bar {\tilde  f}~,~ \chi \chi ~,
H^+H^-, H^0H^{'0}$.
These checks can be done either through a direct measurement
of the polarization of the initial or the final states, whenever
possible; or by looking
at the agreement between the differential
cross section measured experimentally and the theoretical
predictions based on the leading helicity conserving amplitudes.

\vspace{0.5cm}
\noindent{\large\bf{Acknowledgement}}\\
\noindent
G.J.G. gratefully acknowledges also the support by the European Union
RTN contract MRTN-CT-2004-503369.

\newpage

\begin{figure}[b]
\[
\hspace{-0.5cm}\epsfig{file=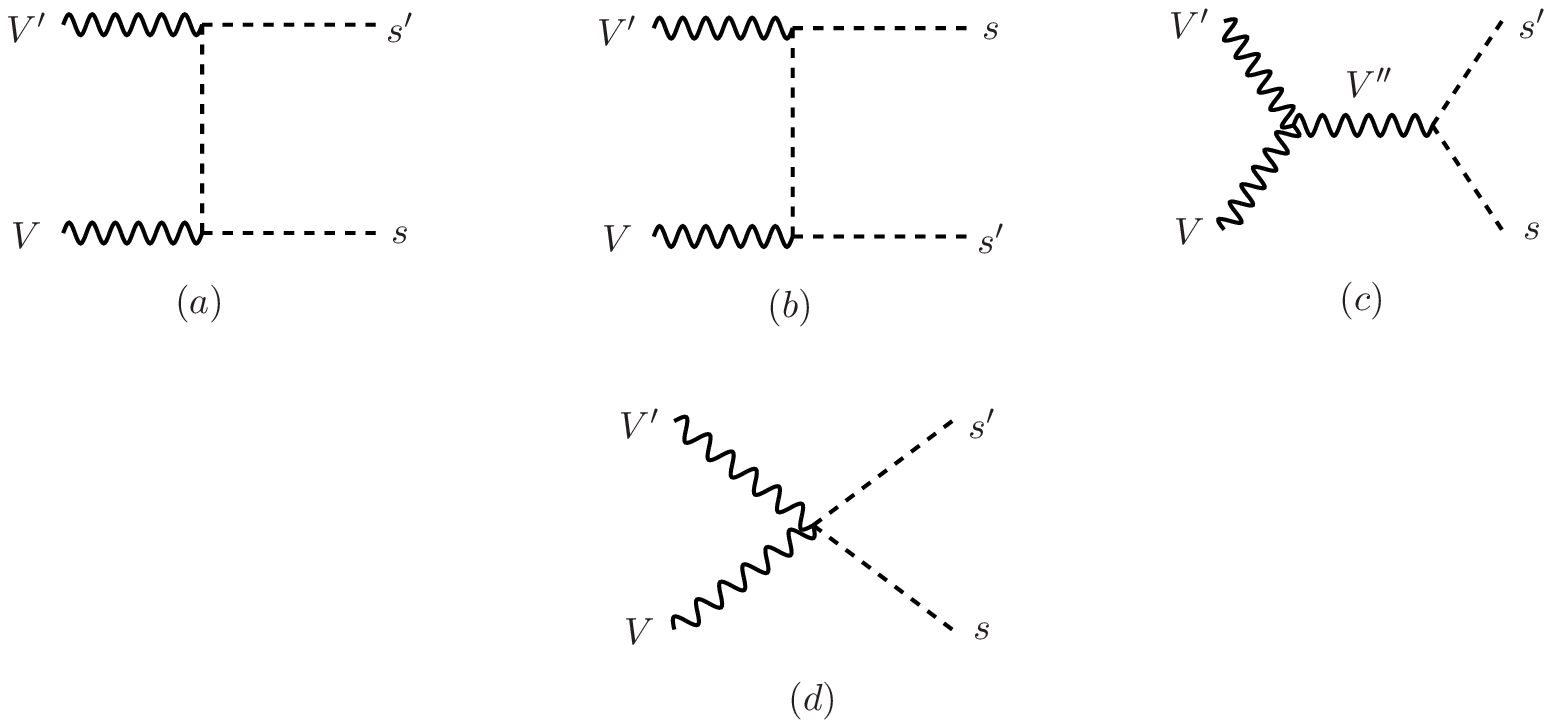,height=6.5cm, width=12cm}
\]
\caption[1]{Born  diagrams for  $VV' \to s s' $,
with $VV'$ being gauge bosons and  $s s'$ being scalar particles.}
\label{VVss-fig}
\end{figure}

\begin{figure}[t]
\[
\hspace{-0.5cm}\epsfig{file=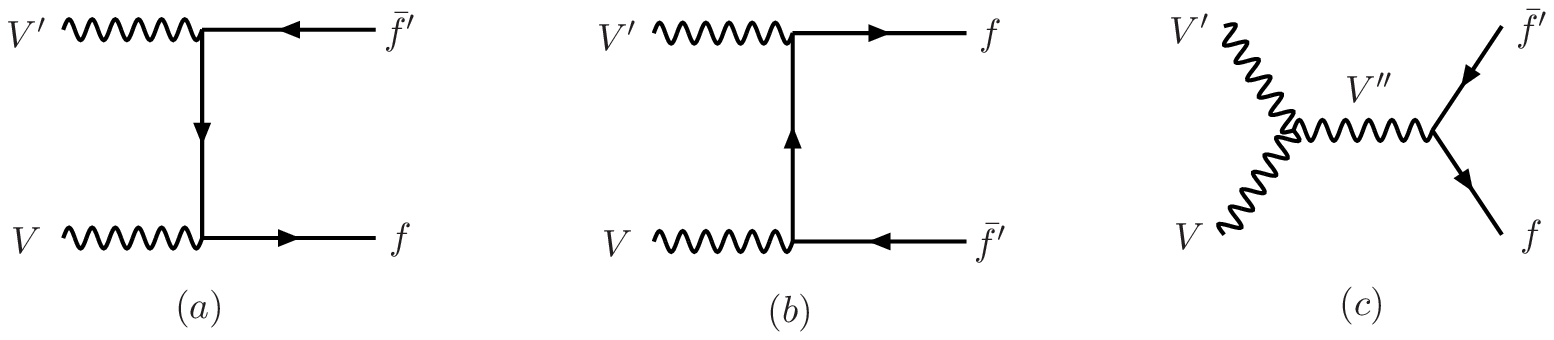,height=3.0cm, width=12cm}
\]
\caption[1]{Born  diagrams for  $VV' \to f \bar f' $ with
$VV'$ being gauge bosons and $f \bar f' $ being fermions.  }
\label{VVff-fig}
\end{figure}


\newpage

\begin{thebibliography}{99}
%
\bibitem{GLRWW} G.J. Gounaris, J. Layssac
and F.M. Renard, \zp{C62}{139}{1994},
arXiv:hep-ph/9309324.
%
\bibitem{Schi} M. Bilenky, J.L. Kneur ,F.M. Renard and D. Schildknecht,
\np{B409}{22}{1993}.
%
\bibitem{GR} G.J. Gounaris and F.M. Renard, \zp{C59}{133}{1993}.
%
\bibitem{BRV} M. Beccaria, F.M. Renard and C. Verzegnassi,
Linear Collider note LC-TH-2002-005;
M. Beccaria, M. Melles, F.M. Renard and C. Verzegnassi,
Phys.Rev.{\bf D65}, 093007 (2002);
M. Beccaria, M. Melles, F.M. Renard, S. Trimarchi and C. Verzegnassi,
arXiv:hep-ph/0304110, IJMP {\bf A18}:5069 (2003);
M. Beccaria, F.M. Renard and C. Verzegnassi,
Phys.Rev.D69,113004(2004).
%
\bibitem{DP} A. Denner, S. Pozzorini, \epj{C18}{461}{2001};
\epj{C21}{63}{2001}.
%
\bibitem{QCD} M. Melles, Phys.Rept. {\bf 375},219(2003).
%
\bibitem{LC-VV} G.J. Gounaris, J.Layssac, and F.M.Renard,
\pr{D67}{013012}{2003}, arXiv:hep-ph/0211327; arXiv:hep-ph/0207273.
%
\bibitem{WW}
M. Beccaria, F.M. Renard and C. Verzegnassi,
Nucl.Phys. {\bf B663},394(2003).
%
\bibitem{Laz-4gamma} G.J. Gounaris, P.I. Porfyriadis
  and F.M.Renard, \epj{C9}{673}{1999}, arXiv:hep-ph/9902230.
%
\bibitem{Laz-ZZ} G.J. Gounaris, J.Layssac, P.I. Porfyriadis
  and F.M.Renard, \epj{C13}{79}{2000}, arXiv:hep-ph/9909243;
  G.J. Gounaris, P.I. Porfyriadis
  and F.M.Renard, \epj{C19}{57}{2001}, arXiv:hep-ph/00100006.
%
\bibitem{Laz-gamZ} G.J. Gounaris, J.Layssac, P.I. Porfyriadis
  and F.M.Renard, \epj{C10}{499}{1999}, arXiv:hep-ph/9904450.
%
\bibitem{SUSY} H. Nilles, Phys.Rep. {\bf 110},1(1984);
H.E. Haber and G.L. Kane, Phys.Rep. {\bf 117},75(1985).
%
\bibitem{equiv} J.M. Cornwall, D.N. Levin, and G. Tiktopoulos,
\pr{D10}{1145}{1974};
M.S. Chanowitz, and M.K. Gaillard, \np{B261}{379}{1985};
G.J. Gounaris, R. K\"ogerler and H. Neufeld, \pr{D34}{3257}{1986};
H. Veltman, \pr{D41}{2294}{1990}.
%
\bibitem{Parke}G. Mahlon and S. Parke \pl{411}{173}{1997}.
%
\end{thebibliography}
\end{document}